\documentstyle[epsfig,12pt,preprint,tighten,aps]{revtex}
\begin{document}
%\input psfig.sty
%\draft

\title{
\rightline{{\tt August 2000}}
%\vskip 1cm
%\rightline{{\tt UM-P-99/xx}}
%\rightline{{\tt RCHEP-99/xx}}
%\rightline{{\tt INFN-ROMA1-xxx/99}}
\ \\
Active-Sterile neutrino oscillations and BBN+CMBR constraints}
\author{P. Di Bari$^{1,2}$ and R. Foot$^2$}
%\date{\today}
\maketitle
%\address{
%\vspace{-1cm}
\begin{center}
{\em
$^1$ Istituto Nazionale di Fisica Nucleare (INFN)\\
$^2$ School of Physics \\
Research Centre for High Energy Physics\\
The University of Melbourne\\
Victoria 3010 Australia\\
(dibari,foot@physics.unimelb.edu.au)}
\end{center}

\begin{abstract}
We show how active-sterile neutrino oscillations in the early Universe
can play an interesting role in explaining the current observations
of CMBR anisotropies and light element abundances. We describe different 
possible phenomenological scenarios in the interpretation of present data 
and how active-sterile neutrino oscillations can provide
a viable theoretical framework.
\end{abstract}
\newpage

\section{Introduction}

The standard big bang model is a simple and testable
theory of the evolution of the Universe.
One of the quantitative tests of
standard big bang cosmology
lies in its predictions of the primordial abundance of light elements.
Standard Big Bang Nucleosynthesis (SBBN)
contains essentially just one free parameter, 
the baryon to photon ratio at the time of BBN $\eta$,
while it predicts the primordial nuclear abundances of several light
elements. 
Each measurement of a primordial abundance, that has to be inferred from
observations at the present time, provides in principle a
measurement of $\eta$. 
The success of the theory relies on the consistency  
of the values for $\eta$ that can be deduced 
from different nuclear abundances. 

%Serious hints suggesting that SBBN may need to be
%modified in some way have been emerged during the last few years.
%These hints were basically due to a discrepancy between the value
%of $\eta$ inferred from  the measurement of $^4$He 
%abundance with that from the Deuterium abundance. 

In recent years two different measurements of the
Helium abundance have been given. A first group \cite{Steigman} finds 
\footnote{Where not otherwise indicated, all errors are  meant at 68\% c.l.}
`low' values $Y_p=0.234\pm 0.003$,
while a second group \cite{Izotov} finds `high' values 
$Y_p=0.244\pm 0.002$, which are mutually compatible
at about $2.5\,\sigma$ level only.

The SBBN provides numerically a relation $Y_p(\eta)$.
A linear expansion gives\cite{walker}
\footnote{A neutron life time $\tau_n=887 sec$ 
has been used. Here and everywhere $\eta$ is expressed 
in unit of $10^{-10}$. The central value for $\eta=5$
has been updated according more recent analysis 
\cite{LopezT}. Note that this expression is accurate to within 0.001 
for $3<\eta<10$ while is less accurate for $\eta \lesssim 3$.}:
\begin{equation}\label{eq:YSBBN}
Y_p^{SBBN}(\eta)=0.2467+0.01\,\ln\left({\eta\over 5}\right).
\end{equation}
In this way for the low value of $Y_p$  
one obtains $\eta_{SBBN}^{^{4}He}=1.5\pm 0.4$, 
while for the high value of $Y_p$
one obtains $\eta_{SBBN}^{^{4}He}=3.9 \pm 0.8$.

Meanwhile, the value
$(D/H)_5=3.39\pm 0.25$ 
\footnote{$(D/H)_5=10^5\,(D/H)$.}
has been deduced \cite{Burles}  for the primordial deuterium abundance
from observations toward two high redshift quasars. 
Deuterium is an ideal
`baryometer' \cite{Schramm} which gives an accurate
measurement of baryon abundance in the context of SBBN.
Indeed one finds
$(\Omega_b\,h^2)_{SBBN}^{D/H}=0.019\pm 0.0024$ (95\% cl) \cite{BurlesTurner}
and from the simple relation 
$\eta\simeq 273\,\Omega_b\,h^2$, this corresponds to 
$\eta_{SBBN}^{D/H}=5.2\pm 0.65$ (95\% cl)
\footnote{In a very recent analysis even smaller errors are found:
$(\Omega_b\,h^2)_{SBBN}^{D/H}=0.019\pm 0.0018$ (95\% cl) and correspondingly
$\eta_{SBBN}^{D/H}=5.2\pm 0.5$ (95\% cl)\cite{BurlesTurner2}.}. 
This value is clearly not consistent
with the $\eta$ value obtained from low Helium values 
although it is consistent with the $\eta$ value from the high Helium value.

On a new front, two balloon
experiments provided the first accurate measurements of
acoustic peaks in the Cosmic Microwave Background Radiation (CMBR) anisotropies 
\cite{Boom,Max} and from these observations it has been possible to
infer a value for the baryon to photon ratio.
The BOOMERanG experiment finds 
$(\Omega_b\,h^2)_{CBR}=0.036^{+0.006}_{-0.005}$ \cite{Boom2} while the MAXIMA 
experiment finds $(\Omega_b\,h^2)_{CBR}=0.031^{+0.007}_{-0.006}$ \cite{Max2}.
These two independent measurements are in quite good agreement
and seem to exclude the presence of large systematic errors.  
A combined analysis of the two gives the 
result $(\Omega_b\,h^2)_{CBR}=0.033\pm 0.005$ \cite{BoomMax} that
corresponds to $\eta_{CBR}=9.0\pm 1.4$. 
This value is higher than the BBN predictions (given 
above) from the inferred values from
both Deuterium and Helium. 
%(with the Deuterium value slightly higher
%than the preferred Helium value). 
These discrepancies 
may be due to
systematic errors but it is also
interesting to consider possible explanations in terms
of non standard physics.
One possibility
is that BBN and CMBR are probing different quantities,
as they involve different physical mechanisms and at different times
(see for example \cite{Turner}). Instead we will consider 
this discrepancy as a hint for non standard BBN.

We will consider two view points:
\vskip 0.3cm
\noindent
1) The discrepancy is between CMBR and Helium
while the discrepancy between CMBR and Deuterium is 
due to systematic uncertainties.
This is plausible because
it is very difficult to identify `clean' 
absorption systems providing reliable 
measurements for deuterium and the quoted results that we used
were derived from only two such measurements.
\vskip 0.3cm
\noindent
2) The discrepancy between deuterium and CMBR as well
as the discrepancy between Helium and CMBR are both real
and due to non standard physics
\footnote{Note that we will neglect from our analysis measurements
of $^{7}$Li abundance. This because in recent years various analysis
conclude that there is still a big uncertainty on the level of
depletion of the primordial abundance to the values that 
we currently observe (for a recent review see \cite{Tytler00}).}.

\vskip 0.3cm
\noindent
%In this paper we will consider both of these
%possibilities. 
The main purpose of this paper is to explore the possible
explanations for these discrepancies 
in terms of active-sterile neutrino oscillations in the early
Universe.

At the present time there is very strong evidence for neutrino
oscillations coming from atmospheric, solar and the LSND experiment
(for a review, see e.g. Ref.\cite{lang}).
The atmospheric neutrino anomaly can be solved (most simply)
via approximately maximal $\nu_\mu \to 
\nu_\tau$ oscillations or via $\nu_\mu \to \nu_{s}$ oscillations
\cite{fvy} (where $\nu_s$ represents a hypothetical sterile neutrino).
On the other hand, the observed solar flux
deficit (about $50\%$ of the expected value) suggests
approximately maximal $\nu_e \to \nu_{\mu,\tau}$ 
oscillations or approximately maximal $\nu_e \to \nu_s$ 
oscillations (see e.g. Ref.\cite{f2000} and references there-in).
Finally, the LSND experiment implies the existence of small angle
$\nu_e \to \nu_\mu$ oscillations with $\delta m^2 \sim 1 eV^2$.
The combination of these three neutrino anomalies suggests
the need for at least one sterile neutrino.
Perhaps the most elegant solution to these neutrino 
anomalies poses that each neutrino is maximally mixed
with a sterile partner (with small mixing between the 
generations)\cite{mirror}. Of course, there are many other possibilities.
In any case, for illustrative purposes we will focus
on the simple case  of $\nu_e \to \nu_s$ oscillations in isolation
and discuss some of the other possibilities qualitatively where
appropriate. 

Ordinary-sterile neutrino oscillations have remarkable implications
for the early Universe.  In particular,
ordinary-sterile neutrino oscillations
can generate large neutrino asymmetries in the early 
Universe\cite{ftv,fv1,fv2} (see also Ref.\cite{Khlopov}),
so large in fact as to imply significant modifications to BBN\cite{fv2}.
In the simple scenario of $\nu_e \to \nu_s$ oscillations in isolation, 
large $\nu_e$ asymmetry is directly produced, 
while, in three (or more) meutrino mixing
scenarios an $\alpha$-neutrino asymmetry ($\alpha=\mu,\tau$) could be first
generated by $\nu_{\alpha}\leftrightarrow \nu_s$ oscillations
and then converted into an electron neutrino asymmetry by
a $\nu_{\alpha}-\nu_e$ oscillations \cite{fv2,foot99}.
Focusing on the simple case of direct production, 
we will find the values of mixing parameters 
which might explain the possible discrepancies in SBBN. 
Interestingly, it turns out that the suggested parameter
space implies a $\nu_e$ with mass $\sim 1 eV$ 
which is close to the current experimental bound.
Furthermore it is also consistent with the measurements of the LSND
experiment and thus can be potentially tested in 
the near future at mini-Boone.

The outline of this paper is as follows.
In section II we briefly discuss the relation between
$\Omega_m$ inferred from X-ray measurements in galaxy
clusters and the value suggested by CMBR. Interestingly
both are consistent with $\Omega_m = 1$. In section III
we show how the $\nu_e \to \nu_s$ oscillation generated
$L_{\nu_e}$ can reconcile the high $\eta_{CBR}$ with
the BBN $Y_p$ results (case 1 above).
In section IV we briefly examine ways in which the 
discrepancies between Deuterium, Helium and CMBR may both be
reconciled (case 2 above).
In section V we will present a new
possible phenomenological scenario, in which large scale inhomogeneities
in the nuclear abundances are admitted. Also in this case
active-sterile neutrino oscillations may provide a viable theoretical model.
We conclude in section VI.

\section{Cosmic concordance or discordance?}
In this section we make some comments
on the previous indications of a `baryon catastrophe' from
X-ray measurements in galaxy clusters and how it
might be related to the inferred values of the
baryon number from CMBR observations.
Recall it is possible to estimate the baryon to total
mass ratio in clusters of galaxies from X-ray measurements
\cite{White} obtaining (at 1$\sigma$) that 
$\Omega_b\,h^{3\over 2}/\Omega_m=(0.05\pm 0.01)$ \cite{Fabian}
(this is confirmed also by measurements based on
the SZ effect that give 
$\Omega_b\,h/\Omega_m=(0.06\pm 0.006)$ \cite{Carlstrom}).
If one imposes that $\Omega_{m}=1$ and using a lower limit on $h>0.5$
finds immediately that $\Omega_b\,h^2>0.035$, much bigger than the 
upper bound that is deduced in a SBBN, both from Deuterium and Helium
abundance (but consistent with CMBR
as we will discuss in a moment). 
This `baryon catastrophe' in SBBN was `solved' by assuming
that $\Omega_m$ can be much less than 1 which implies that we need 
to give up the inflationary paradigm ($\Omega_{0}=1$)
or to admit the presence of a large cosmological constant $\Lambda$ 
such that $\Omega_m+\Omega_{\Lambda}=1$. In this way, using
the SBBN value, previously given, for $\Omega_b\,h^2$ from Deuterium abundance, 
one can infer a value for $\Omega_m=0.45\pm 0.15$. This picture
has been supported by the discovery of an acceleration expansion
from SNe Ia \cite{Perlmutter}, that also points to the existence of a 
large cosmological constant term, $\Omega_{\Lambda}$\cite{PerlTurner}. Roughly
these measurements provide the constraint, in the $\Omega_m-\Omega_{\Lambda}$
plane, 
\begin{equation}
\Omega_{\Lambda}=1.3\,\Omega_m+0.4\pm 0.2.
\label{sem}
\end{equation}
After the SN results, 
the first accurate CMBR measurements of the first acoustic peak position
seem to 
confirm the idea of a flat Universe \cite{someone}. 
In this case one immediately deduces
from Eq.(\ref{sem}) a value $\Omega_m=0.25\pm 0.1$
in very good agreement with galaxy cluster measurements when the
SBBN value for $\Omega_b\,h^2$ is assumed: three independent
methods match each other, with good accuracy,  
in a region around the point $(0.3,0.7)$, in the plane 
$(\Omega_m,\Omega_{\Lambda})$ (`cosmic concordance' \cite{Turner2}).

It is clear however that if one now uses the new CMBR estimation
for $\Omega_b\,h^2$, then from galaxy clusters one obtains 
that $\Omega_m=0.8\pm 0.3$ suggesting `cosmic discordance'
(albeit only mildly)
between galaxy cluster measurements and SNe type Ia.
Moreover now the existence of a cosmological constant is not required
any more from galaxy clusters \cite{Lesgourgues}.
This seems to be suggested also from a recent analysis of CMBR data when a large neutrino asymmetry is allowed \cite{Sarkar,Lesgourgues}. 
Thus, overall things are not so clear at the moment.
It has also been argued in Ref.\cite{reiss} that the 
Supernovae evidence that the expansion of the 
Universe is accelerating is not yet compelling.
Whether or not a large cosmological constant exists needs 
to be confirmed independently, perhaps by future
analysis from the planned satellite experiments, MAP and PLANCK
will help.
In the meantime, for people with the theoretical prejudice that 
$\Omega_{\Lambda}$ is negligible, there is now
some good news since X-ray measurements from galaxy clusters 
and the CMBR results are both consistent with $\Omega_m = 1$.

\section{CMBR anisotropies and Helium observations}

Even assuming the high value for the Helium abundance, the resulting
value of $\eta$ and that one deduced from CMBR differ at about 
3$\sigma$ level. 
%Note that at 1 $\sigma$ level CMBR gives a lower limit $\eta>7.6$ 
% that would correspond to  have $Y_p\simeq 0.25$ in SBBN.

Let us consider the situation from a formal point of view that
will be useful for further developments. In the SBBN picture 
the experimental constraint $Y_p^{SBBN}(\eta)=Y_p^{exp}$ 
gives a measurement of $\eta=\eta_{SBBN}^{^4 He}$, as we observed
in the introduction. 
However CMBR gives an independent measurement of $\eta$ and,
assuming it to be a reliable one, provides a simple test for SBBN
as now one has to satisfy the 
constraint $Y_{p}^{SBBN}(\eta_{CBR})=Y_{p}^{exp}$.
Present measurements do not pass this test
and thus SBBN is somewhat discrepant with Helium and 
CMBR observations. 

Taking this as a hint for new physics, it
suggests that we need
to  modify SBBN introducing
a new parameter $X$. 
It is clear that, if in the physical ranges of values for $X$, all
values of $Y_p$ are possible (with $\eta = \eta_{CBR}$), then it is always 
possible to find a value for $X$ 
satisfying the test $Y_{p}(\eta_{CMBR},X)=Y_{p}^{exp}$ in the non
standard BBN model.

It has been known for a long time that this `game' can be performed allowing 
a modification of the standard particle content
before the BBN epoch  \cite{Shvartsman}.
This modification can be parametrized with the 
(extra) number of (light) neutrino species \cite{SteigmanSchramm}: 
\begin{equation}\label{eq:dNr}
\Delta N_{\nu}^{\rho}=\sum_X\,N_{X}^{\rho}-3, \, {\rm with} \;\;\;\;\;\;
N^{\rho}_X\equiv {120\over 7\,\pi^2} {\rho_{X}+\rho_{\bar{X}}\over T_{\nu}^4},
\end{equation}
($X=\nu_e, \nu_{\mu}, \nu_{\tau}+{\rm new\,particle\,species}$),
where $T_{\nu}\equiv T_{d}\,R_{d}/ R $ is a fiducial temperature of
ideal neutrinos that would instantaneously decouple at 
$T_{d}\gg m_e/2$ (but also $T_d\ll m_{\mu}/2$), 
without sharing any entropy release, 
from electron-positron  annihilations, with photons    	 
\footnote{Note that with this definition, in the standard 
model of particle physics one finds that $\Delta N^{\rho}_{\nu}$ 
is not exactly zero, due to the fact that actually neutrinos are
slightly reheated during electron-positron annihilations (see \cite{Lopez} 
and references therein).}.

With this extra parameter, the SBBN prediction 
for $Y_p$ is modified and approximately the change is given by:
\begin{equation}\label{eq:DYpr}
\Delta Y_p(\eta,\Delta N_{\nu}^{\rho})\simeq
{1\over 6}\,Y_p^{SBBN}(\eta)\,\left(1-{Y_p^{SBBN}(\eta)\over 2}
\right)\,{m_n-m_p\over T_{\rm f}^{n/p}}\,{\Delta N^{\rho}\over N^{\rho}_{\rm st}}\simeq
0.012\,\Delta N_{\nu}^{\rho},
\end{equation}
where $T_{\rm f}^{n/p}\simeq 0.75{\rm MeV}$ is the freezing 
temperature of neutron to proton ratio in the standard case,
while $N^{\rho}_{\rm st}=43/8$ is the number of particle species
and again the last expression has been evaluated for $\eta_{CBR}\simeq 9$.

It has also been known for a long time \cite{Fowler} that 
the standard prediction for the neutron to proton ratio is modified
if one allows the electron neutrino and anti-neutrino 
distributions in momentum space to deviate from the standard case in which 
the thermal equilibrium distributions with zero chemical potentials 
are assumed
\footnote{Electron neutrino and anti-neutrinos distributions are directly 
involved in determining the rates of the $\beta$-reactions
($n+\nu_e\leftrightarrow p+e^-$, $n+e^+\leftrightarrow p+\bar{\nu}_e$) 
responsible for the final value of the neutron to proton ratio together with
the neutron decay.}. In this case an infinite number of new non standard parameters,
the values of the extra numbers of electron neutrinos and antineutrinos in each quantum
state for any momentum, can be virtually introduced.  
However, in realistic models, usually the deviations depend on a finite number 
of parameters.

A particularly simple model \cite{Reeves} is obtained when the distributions
depend only on the neutrino degeneracy $\xi_e$, playing the 
role of a new parameter $X$. 
This has to be generated earlier than the BBN epoch and it is usually
assumed that it is generated also before electron neutrino 
chemical decoupling so that $\xi_e+\bar{\xi}_e=0$. This corresponds to
having a neutrino asymmetry given by:
\begin{equation}
L_{\nu_{e}}\equiv \frac{n_{\nu_e}-n_{\bar{\nu_e}}}{n_{\gamma}}
={\pi^2\over 12\,\zeta (3)}\,\left(\xi_e+{\xi_e^3\over \pi^2}\right).
\end{equation}
In this case the modification of the SBBN 
prediction [see eq.(\ref{eq:YSBBN})] for
$\xi_e\ll 1$ is given by:
\begin{equation}
\Delta\,Y_{p}(\eta,\xi_e)\simeq 
-Y_p^{SBBN}(\eta)\,\left(1-{Y_p^{SBBN}(\eta)\over 2}\right)\,\xi_e
\simeq -0.22\,\xi_e,
\end{equation}
where the last expression has been calculated for $\eta_{CMBR}\simeq 9$.
Thus values $\xi_e\simeq 0.035$ and $\xi_e\simeq 0.08$ can 
easily solve the discrepancy between CMBR observations and high and low 
Helium values respectively.

More generally one can allow a deviation from the standard prediction
for the Helium abundance due both to a modification of the expansion rate 
and to distortions of the electron neutrino and anti-neutrino distributions.
In this case one can distinguish two different contributions to the variation of 
the Helium abundance compared to the standard case:
\begin{equation}
\Delta Y_p= \Delta Y_p^{\rho}+\Delta Y_p^{f_{\nu_e}}.
\end{equation}
Note that this distinction is not ambiguous as it could appear. 
One can in fact always calculate at any instant the 
quantity $\Delta N^{\rho}_{\nu}$ from the eq. (\ref{eq:dNr})
and from that deduce the corresponding value of $\Delta Y_p^{\rho}$ 
defined as the value of $\Delta Y_p$ when the 
standard electron neutrino and antineutrino distributions are assumed. 
Afterwards one can calculate 
$\Delta Y_p^{f_{\nu_e}}\equiv \Delta Y_p-\Delta Y_p^{\rho}$.
It will prove to be convenient to define also a  
{\em total effective number of neutrino species} $\Delta N_{\nu}$ 
that  combines both the effect of a modification of the expansion rate 
and that one due to the distortions of electron neutrino 
and anti-neutrino distributions:
\begin{equation}
\Delta N_{\nu}\equiv \Delta N_{\nu}^{\rho}+\Delta N_{\nu}^{f_{\nu_e}},
\;\;{\rm with}\;\;
\Delta N_{\nu}^{f_{\nu_e}}\equiv {\Delta Y_p^{f_{\nu_e}}\over 0.012}
\end{equation}
With these definitions, the procedure to calculate $\Delta N_{\nu}$ and the
specific contribution $\Delta N_{\nu}^{f_{\nu_e}}$ is particularly simple
if one uses the linear expansion (\ref{eq:DYpr}):
$\Delta N_{\nu}\simeq \Delta Y_p/0.012$ and thus 
$\Delta N^{f_{\nu_e}}_{\nu}\simeq \Delta Y_p/0.012-\Delta N^{\rho}_{\nu}$
\footnote{Of course the choice to describe results in terms of 
$\Delta Y_p$ or of $\Delta N_{\nu}$ is simply a matter of taste. One can say
that in the first case the astrophysical point of view is more emphasized 
than that one of particle physics or vice versa in the second case. In this paper we 
clearly prefer the second one.}. 

In the case of an electron neutrino asymmetry created before
the electron neutrino chemical decoupling ($\Rightarrow \xi_e + 
\bar \xi_e = 0$) 
it is easy to see that 
$\Delta N_{\nu}^{f_{\nu_e}}\simeq-18\,\xi_e$ 
\footnote{We remind that this is valid only for $\xi_e
\ll 1$. Note also
that for large neutrino asymmetries ($\xi_e, \xi_\mu, \xi_\tau 
\stackrel{>}{\sim} 0.5$)
these would also give a non negligible
contribution to $\Delta N_{\nu}^{\rho}$}.
In order to reconcile the discrepancy between the Helium abundance measurements
and the CMBR observations, a negative
$\Delta N_{\nu}$ is required. More precisely, imposing the constraint
$Y_p(\eta_{CBR},\Delta N_{\nu})=Y_p^{\rm exp}$, one finds:
\begin{eqnarray}\label{eq:DeltaN}
\Delta N_{\nu} &=& -1.5 \pm 0.4 \,\; {\rm for} \,\; Y_p^{\rm exp} = 0.234 \pm 0.003, \nonumber \\
\Delta N_{\nu} &=& -0.7 \pm 0.3 \,\; {\rm for} \,\; Y_p^{\rm exp} = 0.244 \pm 0.002. 
\end{eqnarray}

One could simply hypothesis that a large pre-existing
asymmetry exists, however a more
subtle (and testable as we will show) possibility is that light
sterile neutrinos exist. Naively, such neutrinos might be 
expected to lead to a positive $\Delta N_{\nu}$ as they would simply
generate a positive $\Delta N_{\nu}^{\rho}$, however it has
been shown\cite{fv2,bfv,foot99} that due 
to the dynamical generation of $L_{\nu_e}$
by the oscillations themselves, there will be also an important
contribution to $\Delta N_{\nu}^{f_{\nu_e}}$ 
and the total effective $\Delta N_{\nu}$ can
be negative if $L_{\nu_e} > 0$
\footnote{The problem to get a number of effective neutrinos less
than three is not new. It also arises in order to alleviate
the tension between Deuterium measurements and low values of
Helium abundance (`BBN crisis') \cite{Hata}. Also in that case 
the same non standard solutions can be invoked.
Rounding up the usual suspects, we have a ${\rm MeV}\,\tau$ neutrino
decaying prior to the onset of BBN \cite{Kawasaki}, the 
existence of a large electron neutrino asymmetry 
\cite{Langacker} and active-sterile neutrino oscillations
\cite{fv2}, that we are re-considering in this new context.}.

 This contribution cannot
be expressed in terms, for example, of the final asymmetry
by a simple relation as in the case of a pre-existing asymmetry.
This is because the asymmetry is generated, in the interesting cases,
below the chemical decoupling temperature and even below 
the thermal decoupling 
temperature and thus is changing during the time near the 
freezing of the neutron to proton ratio.
Moreover, as the thermal equilibrium assumption is not satisfied anymore,
the electron neutrino distribution will deviate from equilibrium and this 
effect has also to be included. 
Thus the results can be only calculated numerically.

 The sign of $L_{\nu_e}$ cannot be predicted
because it depends on the sign and magnitude of the
initial lepton number asymmetries. For the purposes
of this paper, we assume that it is positive since
we need to generate $\Delta N_{\nu} < 0$.
The simplest example of neutrino oscillation generated
$L_{\nu_e}$  is the direct production of
$L_{\nu_e}$ by $\nu_e \to \nu_{s}$ oscillations.
In this case we can ignore the oscillations involving $\nu_\mu,
\nu_\tau$ provided that either their masses are very
small (so that the largest $|\delta m^2|$ belongs 
to the $\nu_e \to \nu_s$ oscillations and the other
oscillations have $|\delta m^2|$ much less than $1\ eV^2$)
or that they do not mix with the $\nu_e,\nu_s$ (i.e.
the $\nu_e, \nu_s$ decouple from the $\nu_\mu, \nu_\tau$ in the neutrino
mass matrix). In this way the mixing is simply described
by two parameters, the difference of squared eigenstate masses $\delta m^2$
and the mixing angle in vacuum, $\sin^2 2\theta_0$.

%This possibility is most naturally realized in models where
%the $\nu_e$ state is the heaviest state (e.g. inverted mass
%hierarchy). Thus, we will assume that\footnote{
%More than one $\nu_s$ okay, e.g. maximal
%$\nu_e \to \nu_{s'}$ okay...etc}
%\begin{equation}
%m_{\nu_e} \gg m_{\nu_\mu}, m_{\nu_\tau}, m_{\nu_s}
%\end{equation}

In {\bf Figure 1} we solve the quantum kinetic equations for
$\nu_e \to \nu_s$ oscillations for $\sin^2 2\theta_0 = 10^{-8}$
and $\delta m^2/eV^2 = -0.25, -0.5, -1.0, -2.0, -4.0.$
(For details of the numerical procedure 
see Ref.\cite{ropa}). Let us now discuss the behaviour
exhibited in this figure.
As already discussed in detail in previous publications
\cite{ftv,fv1,fv2}
the evolution of lepton number can be separated into
three distinct phases. 
At high temperatures the oscillations are damped
and evolve so that $L^{(e)} \ll \eta$ (where
$L^{(e)} \equiv  2L_{\nu_e} + L_{\nu_\mu}
+ L_{\nu_\tau} + \eta$, and $\eta$ is related to the baryon
asymmetry).
In this region the resonance momentum for neutrino oscillations
is approximately the same as anti-neutrino oscillations.
If $\delta m^2 < 0$ (which means that the 
mass eigenstate which is mainly $\nu_s$ is lighter
than the mass eigenstate which is mainly $\nu_e$) then
at a certain temperature, $T_c$, which is given
roughly by\cite{ftv},
\begin{equation}
T_c \sim 
15\left({-\delta m^2\, \cos2\theta_{0} \over 
\text{eV}^2}\right)^{1 \over 6}\ \text{MeV},
\end{equation}
exponential growth of neutrino asymmetry occurs
(which typically generates a neutrino asymmetry
of order $10^{-5}$, as shown in figure 1).
Taking for definiteness that the $L_{\nu_e}$ is
positive, the anti-neutrino oscillation resonance
moves to very low values of $p/T \sim 0.3$
while the neutrino oscillation resonance moves
to high values $p/T \stackrel{>}{\sim} 10$
(see Ref.\cite{fv2} for a figure illustrating this).
The subsequent evolution of neutrino asymmetries,
which is dominated by adiabatic MSW transitions of
the antineutrinos, follows an approximate $1/T^4$ behaviour
until the resonance has passed through the entire distribution. 
The final asymmetry generated is typically in the 
range $0.23 \stackrel{<}{\sim} L_{\nu_e} \stackrel{<}{\sim}
0.37$\cite{fv2}.
Because the oscillations are dominated by adiabatic
MSW behaviour it is possible to use a relatively
simple and accurate formalism to describe the 
evolution of the system at `low temperatures',
$T \stackrel{<}{\sim} T_c/2$.
In fact, we only need to know the values of
the oscillation resonance momentum at $T \sim T_c/2$.
Previous numerical work has already
shown\cite{fv2} that by $T \sim T_c/2$, neutrino asymmetry
is generated such that $0.2 
\stackrel{<}{\sim} p/T \stackrel{<}{\sim} 0.8$
(the precise value depends on $\sin^2\theta_{0}, 
\delta m^2$).
Furthermore the subsequent evolution is approximately
insensitive to the initial value of $p/T$ in this range.
%(provided, of course, that negligible number
%of sterile neutrinos were produced at high temperature).

For full details of the evolution of $L_{\nu_e}$ and
$\Delta N_{\nu}$ in this model see Ref.\cite{foot99}. 
The evolution of the momentum distribution of electron 
neutrinos is also computed and fed into
a BBN code (that is solved concurrently) 
which allows us to compute $Y_p$ for each choice
of $\delta m^2$ and $\sin^{2} 2\theta_0$\cite{foot99}. 
Particularly simple
results are obtained 
when the constraint 
$\sin^2 2\theta_0\,\sqrt{(|\delta m^2|/{\rm eV}^2)} 
\stackrel{<}{\sim} 2.5 \times 10^{-6}$
is imposed. This corresponds to having $\Delta N_{\nu}^{\rho}
\stackrel{<}{\sim} 0.1$
prior the onset of the neutrino asymmetry generation \cite{dll}.
Moreover, for the interesting values $|\delta m^2| \ll 100\,{\rm eV}^2$,
most of the generated neutrino asymmetry and its associated sterile 
neutrino production, occurs below chemical decoupling so that
$\Delta N_{\nu}^{\rho}$ remains negligible.
In this way the only significant contribution to $\Delta N_{\nu}$ derives 
from the $\Delta N^{f_{\nu_e}}_{\nu}$ part which arises from the 
depletion of the $\bar \nu_e$ states as the MSW resonance
passes creating $L_{\nu_e}$ in the process.
For $\delta m^2 \sim -1 \ eV^2$, the large neutrino asymmetry is
generated provided that
$\sin^2 2\theta_0 \stackrel{>}{\sim} \ few \ \times 10^{-10}$ \cite{fv1,fv2}
(which is essentially the adiabatic condition for this system).
With these two constraints on mixing parameters, 
the resulting $\Delta N_{\nu}$ is practically
independent of $\sin 2\theta_0$ and thus we have a full correspondence 
$\Delta N_{\nu} \leftrightarrow \delta m^2$. The result is 
given in {\bf Figure 2}.  From this figure we can
translate the constraint $Y_p (\eta_{CBR},\Delta N_{\nu}) = Y_p^{\rm exp}$ 
on $\Delta N_{\nu}$ [see Eq.\ref{eq:DeltaN}], into 
a sort of `measurement' of $\delta m^2$, i.e. 
\begin{eqnarray}
\Delta N_{\nu} = -1.5 \pm 0.4, \ \Rightarrow \ \delta m^2 = -2.5 \pm 1.0\ eV^2,
\nonumber \\
\Delta N_{\nu} = -0.7 \pm 0.3, \ \Rightarrow \ \delta m^2 = -0.8 \pm 0.5\ eV^2.
\label{del}
\end{eqnarray}
These values of $\delta m^2$ are interesting from
several points of view.
They imply that $m_{\nu_e} \sim 1 \ eV$ (assuming 
$m_{\nu_s} \ll m_{\nu_e}$) which
is close to the present experimental limit.
Furthermore, if $m_{\nu_e}$ is heavier than the $\nu_\mu$ state
then the LSND $\delta m^2$ , $\delta m^2_{lsnd}$ is, approximately,
the same as the $\delta m^2$ for $\nu_e \to \nu_s$ oscillations.
Thus, if this simple scenario is the cause
of the BBN discrepancy it can be potentially tested in the
near future at mini-Boone.

Above we have discussed things in the model
where $L_{\nu_e}$ is produced directly.
It is also possible to produce $L_{\nu_e}$ 
indirectly. E.g. if $\nu_\tau$ is the heaviest
neutrino and oscillations between $\nu_\tau \to
\nu_{s}$ generate a large $L_{\nu_\tau}$
some of which is transferred to $L_{\nu_e}$
by $\nu_e \to \nu_\tau$ oscillations\cite{fv2,bfv,foot99}.
The indirect mechanism typically generates
a smaller $L_{\nu_e}$ leading to
$\Delta N_{\nu}$ in the range 
$-0.7 \stackrel{<}{\sim} \Delta N_{\nu} \stackrel{<}{\sim} 0$ if
$L_{\nu_e} > 0$. 
Models with three sterile neutrinos
(such as models with mirror neutrinos)
have also been studied\cite{mirror2}. These models can also 
accommodate negative $\Delta N_{\nu}$  
in the range 
$-1.5 \stackrel{<}{\sim} \Delta N_{\nu} \stackrel{<}{\sim} 0$ 
if $L_{\nu_e} > 0$. In fact it is fair to say that
a deviation of $\Delta N_{\nu}$ from zero is a generic
consequence of models with light sterile neutrinos if one
of the active neutrinos has mass in the eV range.

\section{`Just so' BBN ?}

We want now to include the Deuterium observations
in our analysis. In this case the discrepancy between
CMBR and nuclear abundances observations, becomes even more
puzzling. In the SBBN $(D/H)(\eta)\propto \eta^{-1.7}$
\cite{SchrammTurner} and this means that having $\eta_{CBR}\sim 9$
corresponds to $(D/H)_5\simeq 1.5$, a quantity about half the
measured one. One could hope that, within the
model of $\nu_e \leftrightarrow \nu_s$ oscillations with $\Delta N^{\rho}_\nu \ll 1$ 
discussed in the previous section, choosing the values of $\delta m^2$ able to reconcile 
CMBR and Helium observations, it would also be possible 
to satisfy the constraint
$(D/H)(\eta_{CBR},\delta m^2)=(D/H)^{exp}$.  This is however not the case
as the negative values of $\Delta N_{\nu}^{f_{\nu_e}}$ leaves almost unchanged 
the standard value corresponding to $\eta_{CBR}$. 
Thus the only way out is to enlarge the 
space of parameters in the model of BBN. This
can be done allowing also a non zero $\Delta N_{\nu}^{\rho}$. This possibility
has also been studied for a long time \cite{Beaudet} and recently
reproposed in \cite{Lesgourgues} to solve the BBN-CMBR discrepancy. 
In this way the problem is now to find values of $\Delta N^{\rho}_{\nu}$ and $\Delta N_{\nu}^{f_{\nu_e}}$ that satisfy simultaneously the constraints:
\begin{eqnarray}
Y_p(\eta_{CMBR}, \Delta N^{\rho}_{\nu},\Delta N^{f_{\nu_e}}_{\nu}) & = & 
Y^{\rm exp}_p, \\
(D/H)(\eta_{CMBR}, \Delta N^{\rho}_{\nu},\Delta N^{f_{\nu_e}}_{\nu}) & = & 
(D/H)^{\rm exp}.
\end{eqnarray}
In a recent analysis \cite{Esposito2}, in which a pre-existing 
electron neutrino
asymmetry is assumed, the authors find that to reduce the discrepancy 
within a $2\sigma$ level, a range of values 
$1 \stackrel{<}{\sim} \Delta N^{\rho}_{\nu}\stackrel{<}{\sim} 11$ 
and correspondingly
$0.07 \stackrel{<}{\sim} \xi_e \stackrel{<}{\sim} 0.43$ must be chosen. 

As we said in the previous section, it is not possible to 
make a straightforward
comparison with the active-sterile neutrino oscillations, 
as the effect of the generation
of a neutrino asymmetry is not easily related. However we 
can make some qualitative 
comments. It is quite easy to have 
marginal consistency at the $2\sigma$ level
by just modifying the constraints imposed on the mixing parameters to solve
the discrepancy of CMBR with the Helium abundance alone. In fact simply
increasing the mixing angle, with a fixed $\delta m^2$, such that
$\sin^2 2\theta_0\,\sqrt{(|\delta m^2|/{\rm eV}^2)} \stackrel{>}{\sim}
2\times 10^{-5}$,
one gets a $\Delta N^{\rho}_{\nu}\stackrel{>}{\sim} 0.6$. It is likely that
the values for $\delta m^2$ found in the previous section 
will be slightly increased as
a higher $\Delta N^{f_{\nu_e}}_{\nu}$ is now required to satisfy
also the constraint from the Helium abundance
\footnote{It must be said that increasing the mixing angle
there is a region where at the onset of the asymmetry 
generation rapid oscillations
are found \cite{ropa} (see also Ref.\cite{eks}). 
It is still an issue whether this is a real feature if the solutions or
just simply an effect due to numerical inaccuracy. 
However if this effect would really exist,
it is possible that the
sign of the asymmetry could be randomly determined in different
points of the space with the creation of lepton domains \cite{ShiFuller}. 
This would spoil the effect that we want to get, as in this 
case negative values
of $\Delta N_{\nu}$ would not be allowed.
In any case at the high values of mixing angles that we are requiring
in order to have $\Delta N_{\nu}^{\rho}\simeq 1$ 
(of course $\Delta N^{\rho}_{\nu}$ cannot be too close to 
one otherwise this
will suppress the final neutrino asymmetry), there are surely  
no rapid oscillations \cite{ropa}.}. 
Therefore, within the framework of
active-sterile neutrino oscillations, the search for the suitable values
for $\Delta N^{f_{\nu_e}},\Delta N^{\rho}$ is translated in a search for the right 
$\delta m^2,\sin^2 2\theta_0$ values.

It is clear however that allowing for the existence of just 
one sterile neutrino species,
values of $\Delta N^{\rho}_{\nu}\sim 5$, required to have a best fit, 
are not possible. In this case one has necessarily to assume
the existence of more than one light sterile neutrino species.
One amusing possibility is the idea that a mirror world
exists where every particle has a corresponding mirror 
particle\cite{mirror3} (see also Ref.\cite{mirror} and references
there-in).  The main theoretical motivation
for this theory is that it allows parity and time reversal to be exact
unbroken symmetries of nature.
In the context of this theory, it is usually assumed that the temperature of
the mirror particles is less than the ordinary ones in
the early Universe\cite{mirror2,mirror}. However,
it is possible that interactions between the ordinary and mirror
worlds may be strong enough to thermalize the mirror particles
such that $T_{mirror} = T_{ordinary}\equiv T$.
In this case $\Delta N_{\nu}^{\rho} \simeq 6.14$.
To reconcile BBN with such a large value of $\Delta N_{\nu}^{\rho}$
requires a large $\xi_{e} \approx 0.4$ pre-existing asymmetry.
(It needs to be pre-existing because if $T_{mirror} = T$,
then there are equal densities of ordinary and mirror neutrinos
which means that one cannot generate significant asymmetries).
Alternatively, if $0.7\,T < T_{mirror} < T$, then
$1.5 \stackrel{<}{\sim} \Delta N^{\rho}_{\nu}
\stackrel{<}{\sim} 6.14$.
In this case neutrino oscillations can generate a 
significant $\nu_e$ asymmetry
which may potentially lead to a model consistent with BBN
for a range of parameters\footnote{
It may also be possible for $T_{mirror} > T$, with large 
$L_{\nu_e}$ generated by $\nu'_{\tau} \to \nu_e$ oscillations, leading
to a consistent model for a range of parameters.}.

While such possibilities are interesting and testable,
it may however seem surprising that nature should have large
$\Delta N^{\rho}_{\nu}$ and $\Delta N^{f_{\nu_e}}_{\nu}$ which roughly
cancel
\footnote{A model employing a decaying Mev $\tau$ neutrino has also
been recently proposed to get a `just so' BBN scenario \cite{Hansen}.}.
%However it is hard to believe 
%that nature choose such a combination of parameters for which one has two
%large deviations from the SBBN that cancel each other yielding 
%values for the nuclear abundances compatible with a SBBN. In other words 
%in last thirty years we essentially should have 
%trusted a model that appears correct
%but is completely wrong. 
Maybe the discrepancy will be alleviated
from more precise measurements of $\eta$ from CMBR and a mild compensation
with one or two extra neutrino species and a not too big neutrino asymmetry
would be perhaps reasonable. From this point of view a crucial
test in the future will be provided when CMBR will also be able to measure
$N^{\rho}_{\nu}$, while at the moment it only provides a 
rather poor upper limit $N_{\nu}^{\rho}\leq 13$ \cite{Hannestad}.
It is however possible to imagine a different kind of solution within
non standard BBN models that circumvent 
the requirement of a fine tuned solution.
We now turn our attention to one idea in this direction.

\section{Inhomogeneous nuclear abundances ?}

CMBR measures the baryon abundance on the whole observable universe
with a comoving size of $6000 {\rm Mpc} \,h^{-1}$ and this
would correspond to a SBBN prediction of $Y_p\simeq 0.25$
and $(D/H)_5\simeq 1.5$. The Deuterium measurement is deduced from
two Lyman absorption systems at $z_{\rm abs}\simeq 3$, corresponding to
comoving distances of about $2000 {\rm Mpc}$. The size of these systems
is approximately equal to the comoving size of galaxies ($100 {\rm Kpc}
- 1 {\rm Mpc}$).
Primordial Helium abundance values are deduced 
from ionized gas surrounding hot young
stars at distances within $\sim100 {\rm Mpc}$ around us. It is then possible to imagine that
an inhomogeneous electron neutrino asymmetry could be the reason for the
apparent discrepancies between Deuterium
and CMBR as well as between Helium and CMBR.
The quantities $Y_p^{SBBN} (\eta_{CBR}) \simeq 0.25, D/H^{SBBN} (\eta_{CBR})
\simeq 1.5$ 
provide us the values
of the nuclear abundances as they would be in absence of neutrino asymmetry.
This would mean that in the absorption systems that we observe, a large
negative neutrino asymmetry is needed to change $(D/H)_5$ from 
$\sim 1.5$ to $\sim 3$. On the other hand to
explain values of $Y_p$ less than $0.25$ in our
surroundings, as we already discussed at length, a
positive neutrino asymmetry is required. Note that a hint of the 
presence of inhomogeneities in Deuterium abundances comes from the observation
of Deuterium in a $z_{\rm abs}=0.701$ toward QSO 1718+4807 where it was found
$(D/H)_5=25\pm 5$ \cite{Webb}. Other authors repeated the analysis 
and, even though they confirm an high value, they arrive at a much looser bound, 
$(D/H)_5=8 - 57$, concluding that the determination of $D/H$ from QSO 1718+4807 
is uncertain \cite{Tytler3}
\footnote{They observe in fact that in this case the spectra of the Lyman series lines 
is missing. This is needed to determine the velocity distribution
of the Hydrogen and these measurements with the high value 
assume a single velocity component.}. 
Using a more elaborate model for the velocity distribution inside 
the absorber, a third group finds $(D/H)_5=4.1 - 4.7$\cite{Levshakov}, 
in any case still higher than the value, $3.3\pm 0.25$ deduced from the two 
cleanest absorption systems.
Another system gives a result $(D/H)_5< 6.7$ \cite{Kirkman}. In a recent review
the possibility of high amplitude inhomogeneities with an equal proportion 
of low values
$(D/H)_5\sim 3$ and high values $(D/H)_5\sim 10$ is 
excluded \cite{Tytler00}. However
it cannot be excluded that rare peaks with $(D/H)_5\sim 10$ are 
present and in any case inhomogeneities with values changing in the 
range $(D/H)_5=1 - 4$ cannot be excluded at the moment.
The possibility for Deuterium abundance inhomogeneities 
has already been explained  with the presence 
of inhomogeneous electron chemical 
potential \cite{DolgovPagel}, an interpretation that 
could be now enforced by CMBR data. 

Active-sterile neutrino oscillations can give rise to an inhomogeneous 
field of electron 
neutrino asymmetry when the presence of 
small inhomogeneities in the baryon number is assumed
\cite{domains}. In this case, the generated neutrino asymmetry can have an 
{\em inverted sign} in points where the baryon number is lower than the average 
value. Large scale inhomogeneities in the electron neutrino asymmetry 
might be expected to generate inhomogeneities in 
the energy densities that would leave an imprint
in the CMBR anisotropies that we do not observe. However in the 
case of active-sterile
neutrino oscillations, inhomogeneities in the 
electron neutrino asymmetry would be
compensated by inhomogeneities in the 
sterile neutrino asymmetry in a way 
that the energy density remains homogeneous and the mechanism is 
not constrained by CMBR.
There is one difficulty however due to the fact that one has 
to require the simultaneous
presence of large scale regions with positive electron 
neutrino asymmetry and negative
neutrino asymmetry. It has been shown \cite{domains} that 
domains with inverted sign  
bigger than $10 {\rm Kpc}$ cannot be generated. In this case even 
though at the onset of BBN one would get
values of abundances in regions with  
positive electron neutrino asymmetry
and also in regions with negative neutrino asymmetry, later on 
astrophysical processes, like
supernovae explosions, would mix the different elements leading to
approximately homogeneous values for the abundances. 

One way to circumvent this is to assume the existence of
two scales. On small scales (less than the diffusion length at the time of freezing of 
neutron to proton ratio, $\sim 100 {\rm pc}$)  baryon number inhomogenities
have to be present. On large scales, as big as required by Deuterium observation, 
the amplitude of these inhomogeneities has to change such that only 
in the regions where it is large enough a structure of small scale lepton domains 
with both signs can form. Neutrino diffusion would afterwards make them merge such that,
in these regions, electron neutrino asymmetry is diluted to negligible values prior 
to the freezing of the neutron to proton ratio. On the contrary in the regions 
where domains did not form, a non zero electron neutrino asymmetry, with 
the {\em normal sign}, would be present \cite{domains}.

In this way one can easily get a field of neutrino asymmetries with values changing between 
zero and some maximum values. This means that it would be easily possible to accommodate 
CMBR with only Deuterium observations (in this case the {\em normal} sign should be positive) 
or with only Helium observations (in this case the {\em normal} sign should be negative).  

If we want to accommodate both Deuterium, Helium and CMBR then we 
need domains with both positive and negative signs on scales larger
than 10 Kpc. However, as we mentioned earlier, this violates the
bound from Ref.\cite{domains}. Actually, the conclusion that domains with 
inverted sign on scales larger than $10 {\rm Kpc}$ cannot be obtained relies 
on a simplified assumption for which domains with inverted sign 
cannot merge with each other. This is what would happen in the presence of a simple
spectrum of baryon inhomogeneities with just two characteristic scale lengths as
we just described : one is the scale of small lepton domains 
($\lesssim 100 {\rm pc}$) and one is a scale that modulates 
the amplitude  of baryon inhomogeneities in a way that 
in some regions lepton domains can form and in some others cannot. 
However in a realistic spectrum of baryon fluctuations 
with a random presence of Fourier components things 
can be much different and
one cannot exclude a priori that in some regions, 
small inverted sign lepton domains
can occupy most of the space and they can merge with each other to 
form very large domains with a scale higher than about $100 {\rm Kpc}$,
both with positive and negative sign. These very large scale domains would give rise to inhomogeneities in the nuclear abundances that could not be washed out by astrophysical processes and would survive until the present. 
\footnote{We have to mention that in another alternative model proposed in \cite{DolgovPagel}
the simultaneous presence of regions 
with positive neutrino asymmetry together with regions
with negative neutrino asymmetry is a natural consequence. 
Here of course we are concentrating our attention on active-sterile neutrino oscillations, 
but the consideration that CMBR could be pointing to the presence of large scale inhomogeneities 
in the nuclear abundances has a general validity.}
A clear signature of this mechanism would be the detection of high values of Helium
($\sim 0.30$) in the regions where Deuterium is also measured with a value 
$(D/H)_5 \sim 3$ while if some peaks with $(D/H)_5\sim 10$ 
really exist, here the Helium abundance should be even at level of $Y_p\sim 0.50$ \cite{DolgovPagel}. 
However these measurements at large distances seem, at the present, 
to be quite challenging. 
Anyway when more measurements from Lyman absorption systems will be available, a clear 
signature of inhomogeneities could be possible. On the other hand if observations
will exclude Deuterium abundance inhomogenities in the range $(D/H)_5=1-4$ or larger, 
then an explanation of the BBN-CMBR discrepancy in terms of a spatially variating 
electron neutrino asymmetry, as we are proposing, would be ruled out. 

\section{Conclusions}

We have discussed the discrepancy between the 
inferred baryon number density, $\eta$ from recent 
CMBR measurements and the value inferred from standard
big bang nucleosynthesis.
This discrepancy may be due to some type of 
systematic error or may hint at new physics.
We have explored one possible explanation in terms
of active  - sterile neutrino oscillations.
We focussed on the simplest example to illustrate 
this possibility, and that is the direct production
of $L_{\nu_e}$ by $\nu_e \to \nu_s$ oscillations.
Within the context of this model, we have shown that 
$\delta m^2 \approx -1 \ eV^2$ is required 
to solve the discrepancy between CMBR and Helium measurements
and this would suggest
that the electron neutrino mass is about 1 eV.
This is a particularly
interesting value, since it is right near the boundary of
current experimental measurements.
While we focussed on the largest discrepancy between
$\eta_{SBBN}^{^{4}He}$
and $\eta_{CBR}$, we also discussed the 
$\eta_{SBBN}^{D/H}$
and $\eta_{CBR}$ discrepancy, and its implications
for models with sterile neutrinos. In particular, we looked
at two possible scenarios. The  first one, 
would reconcile the deuterium discrepancy with 
a large $\Delta N^{\rho}_{\nu}$, 
while still needing a $\Delta N^{f_{\nu_e}}_{\nu}$
of the opposite sign to reconcile the Helium measurements.
We also proposed a second scenario in which we argue 
that an inhomogeneous electron neutrino asymmetry
could exist which solves these discrepancies.
For both of them we showed how active-sterile neutrino oscillations
can provide a viable theoretical framework.

Clearly things will soon become more interesting as more accurate
measurements of CMBR and light element abundances are done, and
also, as we learn more about neutrinos from current and
future experiments. Thus, it seems that large neutrino
asymmetries, as generated from active - sterile 
neutrino oscillations offer an exciting interconnection between the rapidly
developing fields of neutrino physics and early Universe cosmology.

\vskip 0.5cm
\noindent
{\bf Acknowledgements}
We thank X-G. He, S. Hansen, M. Lusignoli, R. Volkas and the 
anonymous referee for comments on the paper. We also wish to thank 
G. Steigman who pointed out a wrong quotation, in a earlier 
version, for the value of the predicted primordial  Helium abundance in SBBN.

\newpage
{\bf Figure Captions}
\vskip 0.5cm
\noindent
Figure 1: Evolution of $L^{(e)} = 2L_{\nu_e} + \eta$
for $\nu_e \to \nu_s$ oscillations with $\sin^2 2\theta_0
= 10^{-8}$ and, from left to right, $\delta m^2/eV^2 =
-0.25, -0.5, -1.0, -2.0, -4.0$ obtained
from numerically solving the quantum kinetic
equations. The initial $L_{\nu_e} = 0$
is taken and $\eta = 5\times 10^{-10}$ is assumed.
Of course the low temperature evolution 
is approximately independent of these values.
\vskip 0.5cm
\noindent
Figure 2: Change in the effective number of
neutrinos for BBN, $\Delta N_{\nu}$ versus $-\delta m^2$ for
the case $L_{\nu_e} > 0$.

\newpage
\epsfig{file=fig1.eps,width=15cm}
\newpage
\epsfig{file=fig2.eps,width=15cm}

\begin{thebibliography}{10}

\bibitem{Steigman} 
K.A. Olive and G. Steigman, {\it Astrophys. J. Suppl.} {\bf 97}, 49 (1995).

\bibitem{Izotov} 
Y.I. Izotov and  T.X. Thuan, Ap.J. {\bf 500}, 188 (1998).

\bibitem{walker}
T.P. Walker, G. Steigman, D.N. Schramm, 
K.A. Olive and H.S. Kang, Ap.J. {\bf 376}, 51 (1991).

\bibitem{LopezT}
N. Hata, R.J. Scherrer, G. Steigman, D. Thomas and T.P. Walker,
Ap. J. {\bf 458}, 637 (1996); R. Lopez and M.S. Turner,
Phys. Rev. {\bf D59}, 103502 (1999); K.A. Olive, G. Steigman and T.P. Walker,
Phys.Rept. {\bf 333-334}, 389, (2000).  

\bibitem{Burles}
S. Burles and D. Tytler, Ap.J. {\bf 499}, 699 (1998);
S. Burles and D. Tytler, Ap.J. {\bf 507}, 732 (1998).

\bibitem{Schramm}
H. Reeves, J. Audouze, W.A. Fowler and D.N. Schramm, 
Ap.J. {\bf 179}, 909 (1973).

\bibitem{BurlesTurner}
S. Burles, K.M. Nollett, J.N. Truran and M.S. Turner, 
Phys. Rev. Lett. {\bf 82}, 4176 (1999)

\bibitem{BurlesTurner2} 
S. Burles, K.M. Nollett, and M.S. Turner, astro-ph/0008495.

\bibitem{Boom} P.\ de Bernardis et al., Nature {\bf 404}, 995 (2000),
  astro-ph/0004404.  
  
\bibitem{Max} S.\ Hanany et al., Astrophys.J. Lett.  submitted (2000),
  astro-ph/0005123.

\bibitem{Boom2} A.E.\ Lange et al., Phys. Rev. {\bf D} submitted (2000), astro-ph/0005004

\bibitem{Max2} 
A.\ Balbi et al., Astrophys.J. Lett. submitted (2000),
astro-ph/0005124. 

\bibitem{BoomMax} 
A.H. Jaffe et al., astro-ph/0007333.

\bibitem{Turner}
M.K. Kaplinghat and M.S. Turner, astro-ph/0007454.

\bibitem{Tytler00}
D. Tytler, J. M. O'Meara, N. Suzuki and  D. Lubin, 
Physics Scripta {\bf T85}, 12 (2000) (astro-ph/0001318).

\bibitem{lang}
For a review, see e.g.
P. Langacker, Nucl. Phys. B(Proc. Suppl) {\bf 77}, 241 (1999).

\bibitem{fvy}
R. Foot, R. R. Volkas and O. Yasuda,
Phys. Rev. {\bf D58}, 013006 (1998);
P. Lipari and M. Lusignoli, Phys. Rev. {\bf D58}, 073005 (1998);
N. Fornengo, M. C. Gonzalez-Garcia and J. W. F. Valle,
hep-ph/0002147; R. Foot, hep-ph/0007065. 

\bibitem{f2000}
R. Foot, Phys. Lett. {\bf B483}, 151 (2000).

\bibitem{mirror}
R. Foot and R. R. Volkas, Phys. Rev. {\bf D52}, 6595 (1995);
R. Foot, Mod. Phys. Lett. {\bf A9}, 169 (1994). 
R. Foot, H. Lew and R. R. Volkas, 
Mod. Phys. Lett. {\bf A7}, 2567 (1992). 



\bibitem{ftv}
R. Foot, M. J. Thomson and R. R. Volkas, 
Phys. Rev. {\bf D53}, 5349 (1996).

\bibitem{fv1}
R. Foot and R. R. Volkas, Phys. Rev. {\bf D55}, 5147 (1997).

\bibitem{fv2}
R. Foot and R.R. Volkas, Phys. Rev. {\bf D56}, 6653(1997);
Erratum-ibid, {\bf D59}, 029901 (1999).

\bibitem{Khlopov}
M. Yu. Khlopov and S. T. Petcov, Phys. Lett. {\bf B99}, 117 (1981);
Erratum: Phys. Lett. {\bf B100}, 520 (1981).

\bibitem{foot99}
R. Foot, Phys.Rev. {\bf D61}, 023516 (2000).

\bibitem{White}
S.D.M. White, J.F. Navarro, A.E. Evrard and C.S. Frenk, Nature 
{\bf 366}, 429 (1993). 

\bibitem{Fabian}
S.D.M. White and A.C. Fabian, MNRAS {\bf 273}, 72 (1995);
D.A. Boute and C.R. Canizares, Astrophys. J. {\bf 457}, 565 (1996).

\bibitem{Carlstrom}
J. Carlstrom, Physica Scripta, (1999).

\bibitem{Perlmutter}
A.G. Riess et. al, Astron. J. {\bf 116}, 1009 (1998);
S. Perlmutter et. al, Astrophys. J. {\bf 517}, 565 (1999). 

\bibitem{PerlTurner}
S. Perlmutter, M.S. Turner and M. White, Phys.Rev.Lett. {\bf 83}, 670 (1999). 

\bibitem{someone}
S. Dodelson and L. Knox, Phys.Rev.Lett. {\bf 84}, 3523 (2000);
A. Melchiorri, et. al, Astrophys.J. {\bf 536}, L63 (2000);
P. de Bernardis, Nature {\bf 404}, 955 (2000).

\bibitem{Turner2}
M.S. Turner, Proceedings of Particle Physics and the Universe (Cosmo-98), astro-ph/9904051.

\bibitem{Lesgourgues}
J. Lesgourgues and M. Peloso, astro-ph/0004412.

\bibitem{Sarkar}
J.A.~Adams and S.~Sarkar, preprint OUTP-98-70P and talk presented at
the workshop on {\sl The Physics of Relic Neutrinos}, Trieste,
September 1998; J.~Lesgourgues and S.~Pastor, Phys.Rev. {\bf D60}, 103521 (1999);
W. H. Kinney and A. Riotto, Phys. Rev. Lett. {\bf 83}, 3366 (1999).

\bibitem{reiss}
A. G. Riess, astro-ph/0005229.

\bibitem{Shvartsman}
V.F. Shvartsman, JETP Lett.{\bf 9}, 184 (1969). 

\bibitem{SteigmanSchramm}
G. Steigman, D.N. Schramm and J.E. Gunn, 
Phys. Lett. {\bf B66}, 202 (1977).

\bibitem{Lopez}
R.E. Lopez, S. Dodelson, A. Heckler and M.S. Turner,
Phys.Rev.Lett. {\bf 82}, 3952 (1999). 

\bibitem{Fowler}
R.V. Wagoner, W.A. Fowler and F. Hoyle, Astrophys. J. {\bf 148}, 3 (1967);

\bibitem{Reeves}
H. Reeves, Phys. Rev. {\bf D6}, 3363 (1972).

\bibitem{bfv}
N. F. Bell, R. Foot and R. R. Volkas, Phys. Rev. {\bf D58}, 105010 (1998).

\bibitem{Hata}
N. Hata, R.J. Scherrer, G. Steigman, D. Thomas, T.P. Walker, S. Bludman
and P. Langacker, Phys.Rev.Lett. {\bf 75}, 3977 (1995).

\bibitem{Kawasaki}
See e.g.
M. Kawasaki, P.Kernan, H.S. Kang, R.J. Scherrer, G. Steigman and T.P. Walker,
Nucl. Phys. {\bf B419}, 105 (1994).

\bibitem{Langacker}
N. Hata, G. Steigman, S. Bludman and P. Langacker, Phys.Rev. {\bf D55}, 
540 (1997);
T. Kajino and M. Orito, Nucl. Phys. {\bf A629}, 538c (1998).

\bibitem{dll}
P. Di Bari, P. Lipari and 
M. Lusignoli, Int. Jou. Mod. Phys. {\bf A15}, 2289 (2000).

\bibitem{mirror2}
R. Foot and R. R. Volkas, Astropart. Phys. {\bf A7}, 283 (1997);
Phys. Rev. {\bf D61}, 043507 (2000).

\bibitem{SchrammTurner}
See for example D.N. Schramm and M.S. Turner, 
Rev.Mod.Phys. {\bf 70}, 303 (1998). 

\bibitem{Beaudet}
G. Beaudet and P. Goret, Astron. \& Astrophys. {\bf 49}, 415 (1976);
K. A. Olive, D. N. Schramm, D. Thomas 
and T. P. Walker, Phys. Lett. {\bf B265}, 239 (1991).
H.S. Kang and G. Steigman, Nucl. Phys. {\bf B 372}, 494 (1992).

\bibitem{Esposito2}
S. Esposito, G. Mangano, A. Melchiorri, G. Miele and O. Pisanti, 
astro-ph/0007419.

\bibitem{ropa}
P. Di Bari and R. Foot, Phys. Rev. {\bf D61}, 105012 (2000). 

\bibitem{eks}
K. Enqvist, K. Kainulainen and A. Sorri, Phys. Lett. {\bf B464}, 199 (1999);
X. Shi, Phys. Rev. {\bf D54}, 1753 (1996); R. Buras, hep-ph/0002086.

\bibitem{ShiFuller}
X. Shi and G. Fuller, {\em Phys.Rev.Lett.} {\bf 83}, 3120 (1999).

\bibitem{mirror3}
R. Foot, H. Lew and R. R. Volkas, Phys. Lett. {\bf B272}, 67 (1991).

\bibitem{Hansen}
S.H Hansen and F.L. Villante, Phys.Lett. {\bf B486}, 1 (2000). 

\bibitem{Hannestad}
S. Hannestad, astro-ph/0005018.

\bibitem{Webb}
J.K. Webb, R.F. Carswell, K.M. Lanzetta, R. Ferlet, M. Lemoine,
A. Vidal-Madjar, and D.V. Bowen, Nature {\bf388}, 250 (1997).

\bibitem{Tytler3}
 Tytler, D., {\it et al}., Astron. J. {\bf117}, 63 (1999).

\bibitem{Levshakov}
Levshakov, S.A., Kegel, W.H., \& Takahara, F., A\&A {\bf 336}, L29 (1998).

\bibitem{Kirkman}
D. Kirkman, D. Tytler, S. Burles, D. Lubin, D. \& J. O'Meara, 
Astrophys. J., in press Jan.2000, astro-ph/9907128.

\bibitem{DolgovPagel}
A.D. Dolgov and B.E.J. Pagel, New Astron. {\bf 4}, 223 (1999).

\bibitem{domains}
P. Di Bari, Phys. Lett. {\bf B482}, 150 (2000).



\end{thebibliography}
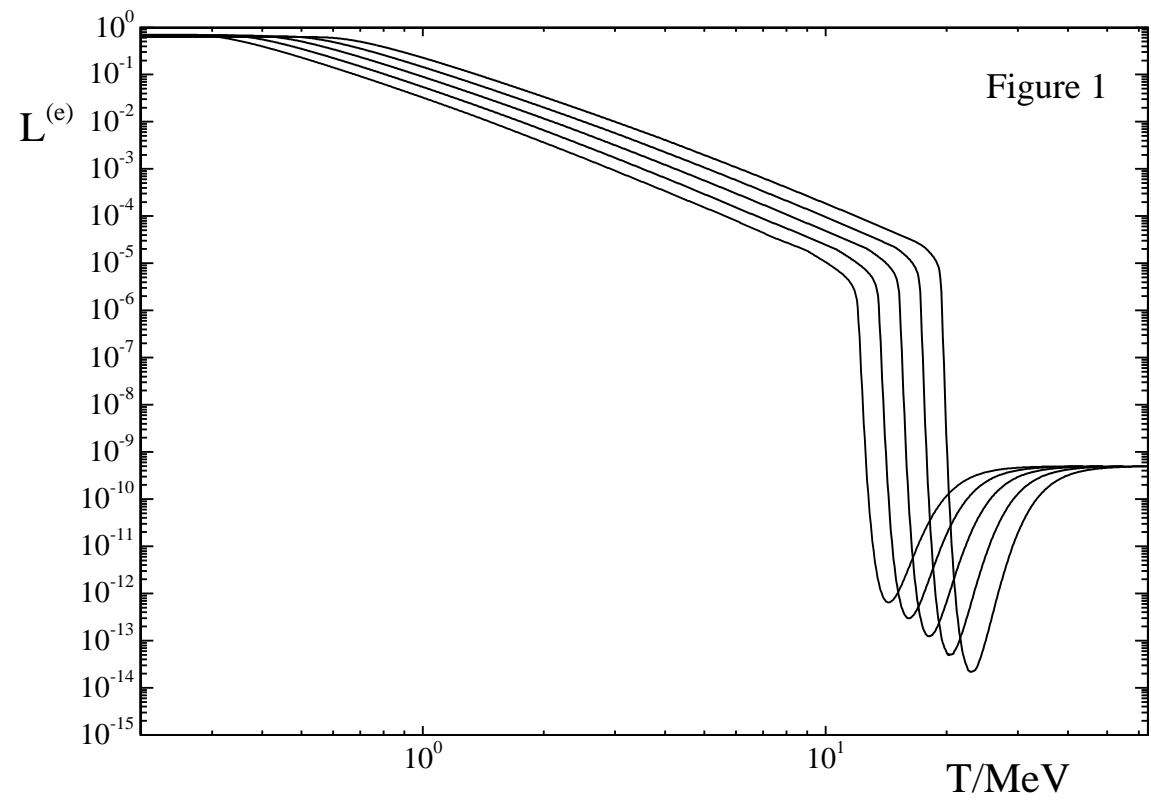
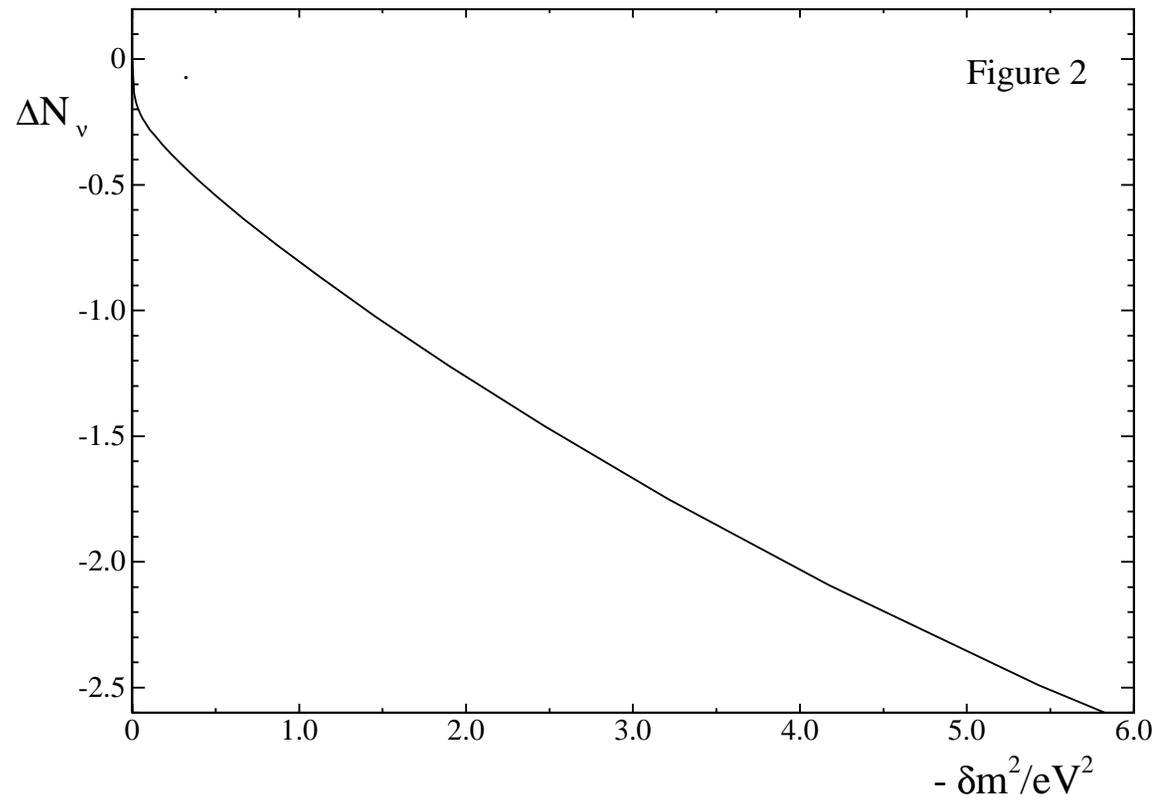
\end{document}